%% file: lcws_arXiv.tex
\documentclass[twoside]{ilcws10}
\usepackage[latin1]{inputenc}
\usepackage[dvips]{graphicx,epsfig,color}
\usepackage{wrapfig,rotating,cite}
\usepackage{amssymb,amsmath,array,axodraw}

\include{paperdef}
\graphicspath{{figs/}}

\begin{document}

\include{lcws_titlepage}

\include{lcws_main}

\end{document}

%% file: lcws_titlepage.tex
\thispagestyle{empty}
\setcounter{page}{0}
\def\thefootnote{\fnsymbol{footnote}}

\begin{flushright}
\mbox{}
DESY 10-112 
\end{flushright}

\vspace{1cm}

\begin{center}

{\large\sc {\bf Top, GigaZ, MegaW}}
\footnote{Invited talk given by S.H.\ at the {\em LCWS/ILC 2010}, 
March 2010, Beijing, China}

\vspace{1cm}

{\sc 
S.~Heinemeyer$^{1}$%
\footnote{
email: Sven.Heinemeyer@cern.ch}%
~and G.~Weiglein$^{2}$%
\footnote{email: Georg.Weiglein@desy.de}
}

\vspace*{1cm}

{\it
$^1$Instituto de F\'isica de Cantabria (CSIC-UC), 
Santander,  Spain\\

\vspace{0.3cm}

$^2$DESY, Notkestra\ss e 85, D--22607 Hamburg, Germany
}
\end{center}

\vspace*{0.2cm}

\BC {\bf Abstract} \EC
\input{lcws_abstract}

\def\thefootnote{\arabic{footnote}}
\setcounter{footnote}{0}

\newpage


%% file: lcws_abstract.tex
We review the physics potential of top mass measurements and the GigaZ/MegaW
options of the International Linear Collider (ILC) for probing New Physics
models and especially the Minimal Supersymmetric Standard Model
(MSSM). 
We demonstrate that the anticipated experimental accuracies at the
ILC for the top-quark mass, $\mt$, the W boson mass, $\MW$, and the effective
leptonic weak mixing angle, $\sweff$, will provide a high
sensitivity to quantum effects of New Physics. In particular, a new and
more precise measurement of $\sweff$, for which the experimental central
value is currently obtained from an average where the most precise
single measurements differ by more than three standard deviations, could
lead to a situation where both the Standard Model and the MSSM in its
most general form are ruled out.
Alternatively, the precision measurements may resolve virtual effects of
SUSY particles even in 
scenarios where the SUSY particles are so heavy that they escape direct
detection at the LHC and the first phase of the ILC.

%% file: lcws_main.tex
\title{Top, GigaZ, MegaW}%
\author{Sven Heinemeyer$^1$, Georg Weiglein$^2$
\vspace{.3cm}\\
1- Instituto de F\'isica de Cantabria (CSIC-UC), Santander, Spain\\
2- DESY, Notkestra\ss e 85, D--22607 Hamburg, Germany \\
}

\maketitle

\begin{abstract}
\input{lcws_abstract}
\end{abstract}


\section{Introduction}

Electroweak precision observables (EWPO) are a very powerful tool for testing 
the Standard Model (SM) and extensions of it. A particularly attractive
extension is the Minimal Supersymmetric Standard Model 
(MSSM), see \citere{PomssmRep} for a review of electroweak
precision physics in the MSSM. In this context the $Z$-pole observables
and $W$~boson physics play an important role. 
The most sensitive observables are
the effective leptonic weak mixing angle, $\sweff$, and the $W$~boson
mass, $\MW$. 
Performing fits in constrained supersymmetric (SUSY) models a certain
preference for not too heavy SUSY particles has been
found~\cite{Master1,Master2,Master3,Master3.5}, see \citere{other}
for other approaches; a detailed list of references can be found in
\citere{Master3}. 
The prospective improvements in the experimental accuracies, in particular
at the ILC with GigaZ option (high luminosity running at the $Z$~pole)
and the MegaW option (high luminosity running at the $WW$~threshold),
will provide a high sensitivity to deviations 
both from the SM and the MSSM.
In \refta{tab:expacc} we summarize the current experimental
results~\cite{LEPEWWG,LEPEWWG2,TEVEWWG} together with the anticipated
improvements at the LHC and the ILC with GigaZ option, see
\citeres{blueband,PomssmRep,gigaz,moenig} for details.

\begin{table}[htb!]
\renewcommand{\arraystretch}{1.5}
\BC
\begin{tabular}{|c||c|c|c|c|}
\hline\hline
observable & central exp.\ value & $\si \equiv \si^{\rm today}$ &
             $\si^{\rm LHC}$ & $\si^{\rm ILC}$ \\ \hline \hline
$\MW$ [GeV] & $80.399$ & $0.023$ & $0.015$  & $0.007$ \\ \hline
$\sweff$    & $0.23153$ & $0.00016$ & $0.00020$--$0.00014$ & $0.000013$ 
                                                      \\ \hline
$\mt$ [GeV]   & $173.3$  & $1.1$    & $1.0$ & $0.1$ \\
\hline\hline
\end{tabular}
\EC
\renewcommand{\arraystretch}{1}
\caption{Summary of the electroweak precision observables, including the
  top-quark mass, 
their current experimental central values and  experimental errors, 
$\si \equiv \si^{\rm today}$~\cite{LEPEWWG,LEPEWWG2,TEVEWWG}. 
Also shown are the anticipated
experimental accuracies at the LHC, $\si^{\rm LHC}$, and the ILC
(including the GigaZ/MegaW options), $\si^{\rm ILC}$. Each number represents
the combined results of all detectors and channels at a given collider,
taking into account correlated systematic uncertainties, see
\citeres{blueband,PomssmRep,gigaz,moenig} for details. 
}
\label{tab:expacc}
\end{table}

The mass of the top quark, $\mt$, is a fundamental parameter of the 
electroweak theory. It is by far the heaviest of all quark masses and
it is also larger than the masses of all other known fundamental
particles. 
It is evident that a
comprehensive program of high-precision measurements at the top
threshold will have to be a key element in the physics program of a
future Linear Collider.
The top quark is deeply connected to many other issues of high-energy
physics: 
\begin{itemize}
\item
The top quark could 
play a special role in/for electroweak symmetry breaking.
\item
The experimental uncertainty of $\mt$ induces the largest parametric
uncertainty in the prediction for electroweak precision
observables~\cite{deltamt} and can thus obscure new physics effects.
\item
In supersymmetric (SUSY) models the top quark mass is an important
input parameter and is crucial for radiative electroweak symmetry
breaking and unification. 
\item
Little Higgs models contain ``heavier tops''.
\end{itemize}
The calculations for $e^+e^- \to t \bar t$ at the 
threshold are quite advanced.
This includes NNLO and NNNLO predictions as well as renormalization
group improved NRQCD calculations, see e.g.\ \citere{topILC} for a review. 
Also for the process
$e^+e^- \to t \bar t H$ and the determination of the top Yukawa coupling 
substantial progress has been made recently, see 
e.g.\,\citere{ttH}.

The large value of $\mt$ gives rise to a large coupling between the top 
quark and the Higgs boson and is furthermore important for flavor
physics. It could therefore provide a window to new physics. (The correct
prediction of $\mt$ will be a crucial test for any fundamental theory.)
The top-quark mass also plays an important role in electroweak precision
physics, as a consequence in particular of non-decoupling effects being
proportional to powers of $\mt$. A precise knowledge of $\mt$ is
therefore indispensable in order to have sensitivity to possible effects
of new physics in electroweak precision tests.

The current world average for the top-quark mass from the measurement
at the Tevatron is 
$\mt = 173.3 \pm 1.1 \gev$~\cite{mt1733}. 
The prospective accuracy at the LHC is 
$\de\mtexp = 1 \gev$~\cite{mtdetLHC}, 
while at the ILC a very precise determination of $\mt$ with an accuracy
of $\de\mtexp \lsim 100 \mev$ will be possible~\cite{lctdrs,mtdet}.
This error contains both the experimental error of the mass parameter
extracted from the $t \bar t$ threshold measurements at the ILC and 
the envisaged theoretical uncertainty from its transition into a suitable
short-distance mass (like the \msbar\ mass).


\section{Top quark mass measurement at the ILC and its implications}

In the following we show for some examples that in many physics 
applications the experimental error on $\mt$ achievable at the LHC
would be the limiting factor, demonstrating the need for the ILC
precision. More examples can be found in \citere{deltamt}.

\subsection{The top quark mass and electroweak precision observables}

In order to confront the predictions of
supersymmetry (SUSY) with the electroweak precision data
and to derive constraints on the supersymmetric parameters,
it is desirable to achieve the same level of accuracy for the
SUSY predictions as for the SM.
In \citeres{MWMSSM,ZObsMSSM} an evaluation of $\MW$ and the $Z$-pole
observables in the MSSM has been presented. It includes the full
one-loop result (for the first time with the full complex phase
dependence), all available MSSM two-loop corrections (entering via the
$\rho$~parameter~\cite{dr2lA,drMSSMal2A,drMSSMal2B}), as well as the
full SM results, see \citeres{MWMSSM, ZObsMSSM} for details. 
The Higgs-boson sector has been implemented including higher-order
corrections (as evaluated with 
{\tt FeynHiggs}~\cite{feynhiggs,mhiggslong,mhiggsAEC,mhcMSSMlong}). 

In addition to the experimental uncertainties, summarized in
\refta{tab:expacc},  there are two sources of
theoretical uncertainties: those from 
unknown higher-order corrections (``intrinsic'' theoretical
uncertainties), and those from experimental errors of the input
parameters (``parametric'' theoretical uncertainties). 
The current and estimated future intrinsic uncertainties within the SM
are~\cite{blueband,mwsweff}
\BEA
\De\MW^{\rm intr,today,SM} \approx 4 \mev, & \quad &
\De\sweff^{\rm intr,today,SM} \approx 5 \times 10^{-5} ~, \\
\De\MW^{\rm intr,future,SM} \approx 2 \mev, & \quad &
\De\sweff^{\rm intr,future,SM} \approx 2 \times 10^{-5} ~,
\label{eq:intruncSM}
\EEA
while in the MSSM the current intrinsic uncertainties are estimated
to~\cite{PomssmRep,drMSSMal2A,drMSSMal2B}
\BE
\De\MW^{\rm intr,today,MSSM} \approx (5 - 9) \mev, \quad
\De\sweff^{\rm intr,today,MSSM} \approx (5 - 7) \times 10^{-5} ~,
\label{eq:intruncMSSM}
\end{equation}
depending on the supersymmetric (SUSY) mass scale. 
In the future one expects that they
can be brought down to the level of the SM, see \refeq{eq:intruncSM}. 

The parametric errors of $\MW$ and $\sweff$ induced by the top quark
mass, the uncertainty of $\De\al_{\rm had}$ (we assume a future
determination of $\de(\De\al_{\rm had}) = 5 \times 10^{-5}$~\cite{fredl})
and the experimental uncertainty of the $Z$~boson mass, 
$\de\MZ = 2.1 \mev$, are collected in \refta{tab:ewpopara}.

\renewcommand{\arraystretch}{1.5}
\begin{table}[htb!]
\begin{center}
\begin{tabular}{|c||c|c|c|c|c|}
\cline{2-5} \multicolumn{1}{c||}{}
& ~$\de\mt = 1 \gev$~ 
& ~$\de\mt = 0.1 \gev$~ & 
~$\de(\De\al_{\rm had})$~ & ~$\de\MZ$~ \\
\hline\hline
~$\de\sweff$ [$10^{-5}$]~ & $3$ & $0.3$ & $1.8$ & $1.4$ \\
\hline
$\De\MW$ [MeV] & $6$ & $1$ & $1$ & $2.5$ \\
\hline\hline
\end{tabular}
\caption{Parametric errors on the prediction of $\MW$ and $\sweff$.}
\label{tab:ewpopara}
\end{center}
\end{table}
\renewcommand{\arraystretch}{1}

In order to keep the theoretical uncertainty induced by $\mt$ at a
level comparable to or smaller than the other parametric and intrinsic
uncertainties, $\de\mt$ has to be of \order{0.1 \gev} in the case of
$\MW$, and about $0.5 \gev$ in the case of $\sweff$. If the
experimental error of $\mt$ remains substantially larger, it
would constitute the limiting factor of the theoretical uncertainty.
Using the EWPO to distinguish different models from each other or to
determine indirectly the unknown model parameters, the ILC precision on
$\mt$ is crucial, in particular in view of the precision measurement
of $\sweff$ at GigaZ~\cite{deltamt}.


\subsection{Top quark mass measurement and Higgs physics}

\begin{figure}[htb]
\BC
\setlength{\unitlength}{1pt}
\begin{picture}(170, 70)
\DashLine(10,40)(60,40){5}
\Text(20,60)[]{$H$}
\Vertex(60,40){3}
\ArrowArc(85,40)(25,0,180)
\ArrowArc(85,40)(25,180,360)
\Text(90,75)[]{$t$}
\Text(90,5)[]{$\bar t$}
\Vertex(110,40){3}
\DashLine(110,40)(160,40){5}
\Text(150,60)[]{$H$}
\end{picture}
\EC
\vspace{-2em}
\caption{Loop contribution of the top quark to the Higgs boson mass.}
\label{fig:mhiggs}
\end{figure}
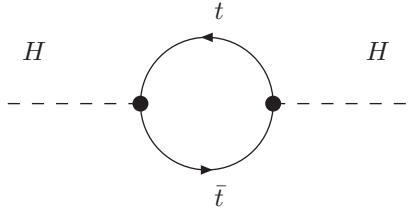

Because of its large mass, the top quark is expected to have a large
Yukawa coupling to Higgs bosons, being proportional to $\mt$.
In each model where the Higgs boson mass is not a free
parameter but predicted in terms of the the other model parameters
(as e.g.\ in the MSSM), the diagram in \reffi{fig:mhiggs} contributes
to the Higgs mass. This diagram gives rise to a leading $\mt$
contribution of the form
\BE
\De\mh^2 \sim \gf \; N_C \; C \; \mt^4~,
\end{equation}
where $\gf$ is the Fermi constant, $N_C$ is the color factor, and the
coefficient $C$ depends on the specific model. Thus the experimental
error of $\mt$ necessarily leads to a parametric error in the Higgs
boson mass evaluation. 

Taking the MSSM as a specific example 
(including also the scalar top contributions and the appropriate
renormalization) $N_C \, C$ is given for the light $\cp$-even Higgs
boson mass in leading logarithmic approximation by  
\BE
N_C \, C = \frac{3}{\sqrt{2}\,\pi^2\,\SQb} \; 
\log \KL \frac{\mste\mstz}{\mt^2} \KR~.
\end{equation}
Here $m_{\tilde t_{1,2}}$ denote the two masses of the scalar tops.
The optimistic LHC
precision of $\de\mt = 1 \gev$ leads to an uncertainty of 
$\sim 2.5\%$ in the prediction of $\mh$, while the ILC will yield a
precision of $\sim 0.2\%$.  
These uncertainties have to be compared with the anticipated precision of
the future Higgs boson mass measurements. With a precision of 
$\de\mh^{\rm exp,LHC} \approx 0.2 \gev$~\cite{atlastdr,cmstdr} the relative
precision is at the level of $\sim 0.2\%$. It is apparent that only the
ILC precision of $\mt$ will yield a parametric error small enough to
allow a precise comparison of the Higgs boson mass prediction and its
experimental value. 

Another issue that has to be kept in mind here (in SUSY as in any other
model predicting $\mh$) is the intrinsic theoretical uncertainty due to
missing higher-order corrections.
Within the MSSM currently this uncertainty is estimated to 
$\de\mh^{\rm intr,today} 
\approx 3 \gev$~\cite{PomssmRep,mhiggsAEC}%
\footnote{We are not aware of any such estimate in other New Physics
models.}%
. This uncertainty could go down by 
$\sim 1 \gev$ once the recent three-loop corrections obtained in
~\citere{mhiggs3l} will be included. In the
future one can hope for an improvement down to 
$\lsim 0.5 \gev$ or better~\cite{PomssmRep,habilSH}, 
i.e.\ with sufficient effort on higher-order corrections it should be possible
to reduce the intrinsic theoretical uncertainty to the level of 
$\de\mh^{\rm exp, LHC}$. 

Confronting the theoretical prediction of $\mh$ with a precise
measurement of the Higgs boson mass constitutes a very sensitive test
of the MSSM (or any other model that predicts $\mh$), which allows one to
obtain constraints on the model 
parameters. However, the sensitivity of the $\mh$ measurement
cannot directly be translated into a prospective indirect determination
of a single model parameter. In a realistic
situation the anticipated experimental errors of {\em all} relevant SUSY
parameters have to be taken into account.
For examples including these parametric errors see \citeres{deltamt,lhcilc}. 


\section{\boldmath{$\MW$} and \boldmath{$\sweff$} in a global MSSM scan}

\begin{figure}[htb!]
\begin{center}
\includegraphics[width=0.65\textwidth,height=10cm]{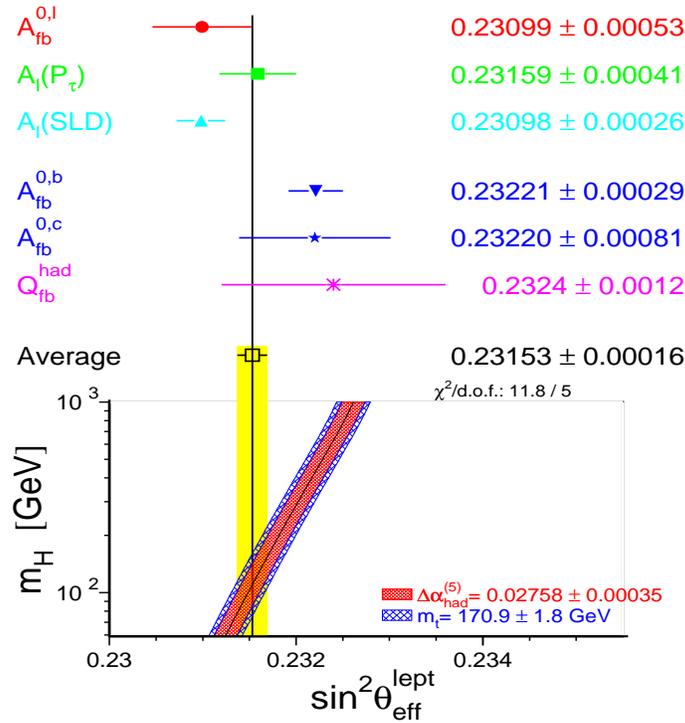}
\end{center}
\vspace{-2em}
\caption{
Individual measurements and world-average of 
$\sweff$. The experimental results are compared with the prediction within the
SM as a function of $\MHSM$ for $\mt = 170.9 \pm 1.8 \gev$ and 
$\De\al_{\rm had}^5 = 0.02758 \pm 0.00035$~\cite{gruenewald07}. 
}
\label{fig:sw2effSM}
\end{figure}

The effective weak mixing angle is determined from various asymmetries
and other EWPO as shown in \reffi{fig:sw2effSM}~\cite{gruenewald07}. 
The world average for the effective weak mixing angle is
\begin{align}
\de\sweff^{\rm exp} &= 0.23153 \pm 0.00016~,
\label{sweff-exp}
\end{align}
with a
$\chi^2/{\rm d.o.f}$ of $11.8/5$, corresponding to a probability of
$3.7\%$~\cite{LEPEWWG,gruenewald07}%
. The large $\chi^2$ is driven by the two
single most precise measurements, $A_{\rm LR}^e$ by SLD and 
$A_{\rm FB}^b$ by LEP, corresponding to
\begin{align}
\label{afb}
A_{\rm FB}^b({\rm LEP}) &: \sweff^{\rm exp,LEP} = 0.23221 \pm 0.00029~, \\
\label{alr}
A_{\rm LR}^e({\rm SLD}) &: \sweff^{\rm exp,SLD} = 0.23098 \pm 0.00026~.
\end{align}
The earlier (latter) one prefers a 
value of $\MHSM \sim 32 (437) \gev$~\cite{gruenewaldpriv}. 
The two measurements differ by more than $3\,\si$.
The averaged value of $\sweff$, as shown in \reffi{fig:sw2effSM},
prefers $\MHSM \sim 110 \gev$~\cite{gruenewaldpriv}.

We now analyse the sensitivity of $\MW$ and $\sweff$ to higher-order
effects in the MSSM by
scanning over a broad range of the SUSY parameter space. The following SUSY
parameters are varied independently of each other in a random parameter scan
within the given range:
\begin{eqnarray}
 {\rm sleptons} &:& M_{{\tilde F},{\tilde F'}}= 100\dots2000\gev, \non \\
 {\rm light~squarks} &:& M_{{\tilde F},{\tilde F'}_{\textup{up/down}}}
                   = 100\dots2000\gev, \non \\
 \Stop/\Sbot {\rm ~doublet} &:& 
                         M_{{\tilde F},{\tilde F'}_{\textup{up/down}}}
                    = 100\dots2000\gev, 
 \quad A_{\tau,t,b} = -2000\dots2000\gev, \non \\
 {\rm gauginos} &:& M_{1,2}=100\dots2000\gev, 
 \quad \mgl=195\dots1500\gev, \non \\
 && \mu = -2000\dots2000\gev,\non \\ 
 {\rm Higgs} &:& \MA=90\dots1000\gev, 
 \quad\tb = 1.1\dots60.
\label{scaninput}
\end{eqnarray}
Here $M_{{\tilde F},{\tilde F'}}$ are the diagonal soft SUSY-breaking
parameters in the sfermion sector, $A_f$ denote the trilinear couplings,
$M_{1,2}$ are the soft SUSY-breaking parameters in the
chargino and neutralino sectors, $\mgl$ is the gluino mass, $\mu$ the Higgs
mixing parameter, $\MA$ the $\cp$-odd Higgs boson mass, and $\tb$ is the
ratio of the two vacuum expectation values.
Only the constraints on the MSSM parameter space
from the LEP Higgs searches~\cite{LEPHiggsMSSM,LEPHiggsSM} and the lower
bounds on the SUSY particle masses from direct searches as given 
in \citere{pdg} were taken into account.
Apart from these constraints no other restrictions on the MSSM parameter
space were made.

\begin{figure}[htb!]
\begin{center}
\includegraphics[width=0.88\textwidth]
                {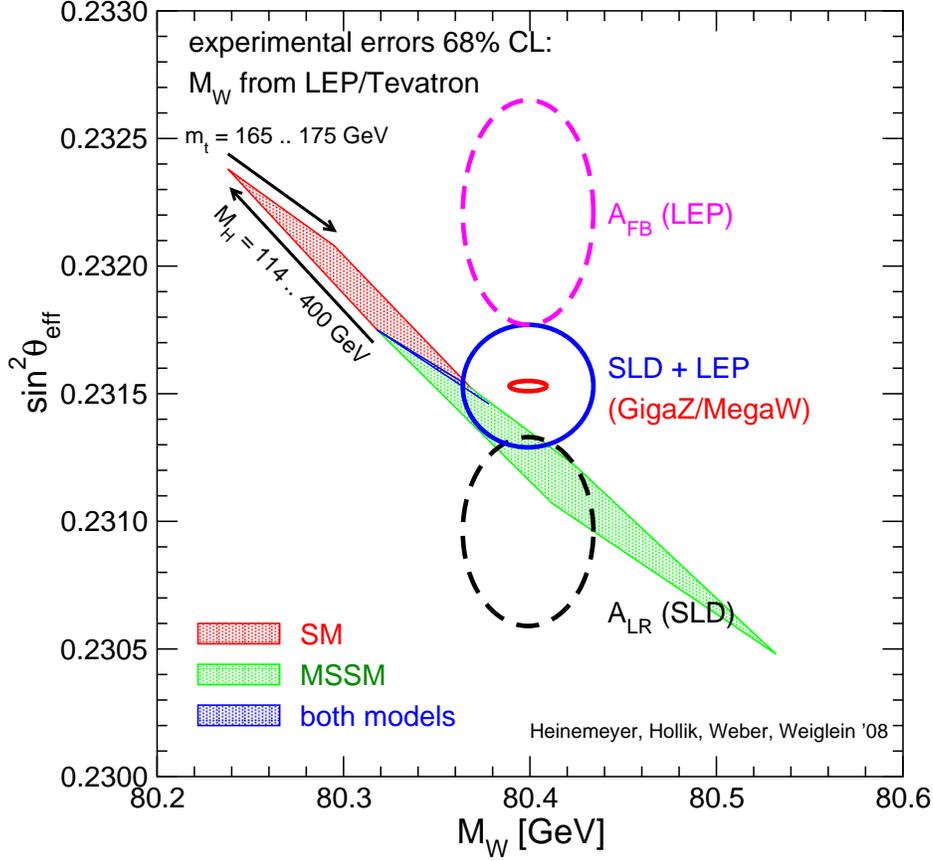}
\end{center}
\vspace{-2em}
\caption{MSSM parameter scan for $\MW$ and $\sweff$ over the
  ranges given in \refeq{scaninput} with 
  $\mt = 165 \ldots 175 \gev$. Todays 68\%~C.L.\ ellipses (from 
  $A_{\rm FB}^b({\rm LEP})$, $A_{\rm LR}^e({\rm SLD})$ and the world
  average) are shown 
  as well as the anticipated GigaZ/MegaW precisions, drawn around todays
  central value}  
\label{fig:Scans2} 
\end{figure}

In \reffi{fig:Scans2} we
compare the SM and the MSSM predictions for $\MW$ and $\sweff$
as obtained from the scatter data. 
The predictions within the two models 
give rise to two bands in the $\MW$--$\sweff$ plane with only a relatively 
small overlap region (indicated by a dark-shaded (blue) area).
The allowed parameter region in the SM (the medium-shaded (red)
and dark-shaded (blue) bands) arises from varying the mass of the SM
Higgs boson, from $\MHSM = 114\gev$, the LEP exclusion bound~\cite{LEPHiggsSM}
(lower edge of the dark-shaded (blue) area), to $400 \gev$ (upper edge of the
medium-shaded (red) area), and from varying $\mt$ in the range of 
$\mt = 165 \ldots 175 \gev$. 
The light shaded (green) and the
dark-shaded (blue) areas indicate allowed regions for the unconstrained
MSSM. The decoupling limit with SUSY masses of \order{2 \tev}
yields the upper edge of the dark-shaded (blue) area. Thus, the overlap 
region between the predictions of the two models corresponds in the SM
to the region where the Higgs boson is light, i.e., in the MSSM allowed
region ($\Mh \lsim 130 \gev$~\cite{mhiggsAEC}). In the MSSM it
corresponds to the case where all 
superpartners are heavy, i.e., the decoupling region of the MSSM.

The 68\%~C.L.\ experimental results
for $\MW$ and $\sweff$ are indicated in the plot. 
The center ellipse corresponds to the current world average given
in \refeq{sweff-exp}. 
Also shown are the error ellipses corresponding to the two individual main
measurements of $\sweff$ as given in \refeqs{afb}, (\ref{alr}).
The anticipated improvement with the GigaZ/MegaW options, indicated as
small ellipse, is shown around the current experimental central data.
One can see that the current averaged value is compatible with the SM
(favoring a light Higgs boson and a heavier top quark) and with the
MSSM. The value of $\sweff$ obtained from $A_{\rm LR}^e$(SLD)
clearly favors the MSSM over the SM.
On the other hand, the value of $\sweff$ obtained from $A^b_{\rm FB}$(LEP) 
together with the $\MW$ data from LEP and the Tevatron would correspond to an
experimentally preferred region that deviates from the predictions of both
models. 
Thus, the unclear experimental situation regarding the two single
most precise measurements entering the combined value for $\sweff$
has a significant impact on the constraints that can be obtained from this
precision observable on possible New Physics scenarios. Measurements at a new 
$e^+e^-$ Z~factory, which could be realized in particular with the GigaZ
option of the ILC, would be needed to resolve this issue.
As indicated by the solid light shaded (red) ellipse, the anticipated
GigaZ/MegaW precision of the combined $\MW$--$\sweff$ measurement could
put severe constraints on each of the models and resolve the discrepancy
between the $A_{\rm FB}^b$(LEP) and $A_{\rm LR}^e$(SLD) measurements. 
If the central value of an improved measurement with higher precision should
turn out to be close to the central value favored by the current measurement
of $A_{\rm FB}^b({\rm LEP})$, this would mean that the electroweak
precision observables $\MW$ and $\sweff$ could rule out both the SM and the
most general version of the MSSM.


\section{Scenario where no SUSY particles are observed  
     at the LHC} 
\label{sec:ILCscen}

It is interesting to investigate whether the high accuracy achievable at
the GigaZ option of the ILC would provide sensitivity to indirect effects of
SUSY particles even in a scenario where the (strongly interacting) 
superpartners are so heavy that they escape detection at the LHC.

\begin{figure}[bth!]
\begin{center}
\includegraphics[width=0.9\textwidth]{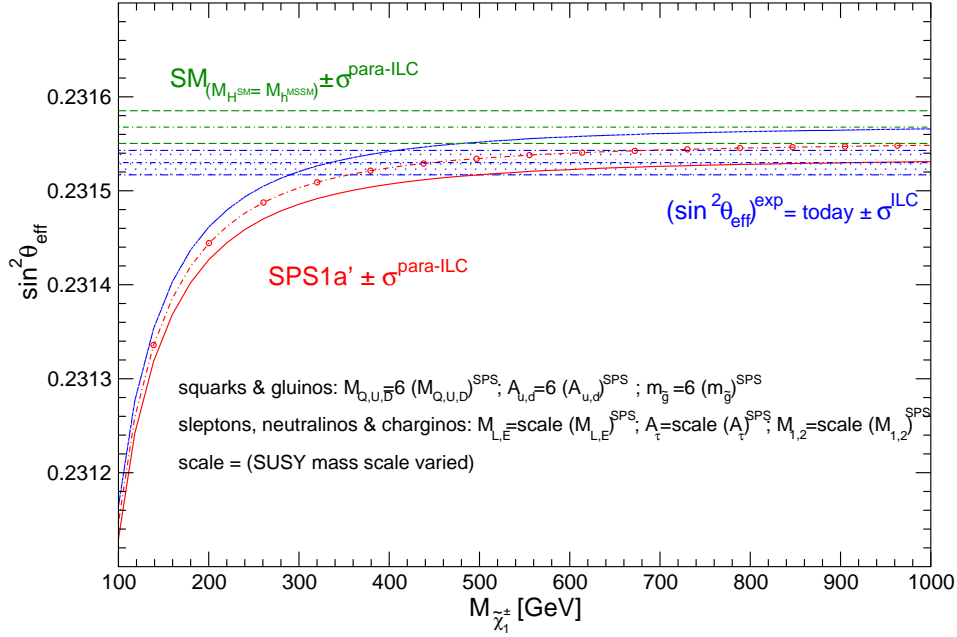}
\vspace{-1.0em}
\caption{
Theoretical prediction for $\sweff$ in the SM and the MSSM (including
prospective parametric theoretical uncertainties) compared to
the experimental precision at the ILC with GigaZ option.  
An SPS1a$'$ inspired scenario is used, where the squark and gluino
mass parameters
are fixed to 6~times their SPS~1a$'$ values. The other mass 
parameters are varied with a common scalefactor.}
\label{fig:ILC} 
\end{center}
\end{figure}

We consider in this context a scenario with very heavy squarks and a 
very heavy gluino. It is based on the values of the SPS~1a$'$ benchmark
scenario~\cite{sps}, but the squark and gluino
mass parameters
are fixed to 6~times their SPS~1a$'$ values. The other masses are 
scaled with a common scale factor 
except $\MA$ which we keep fixed at its SPS~1a$'$ value.
In this scenario 
the strongly interacting particles are too heavy to be detected at the
LHC, while, depending on the scale-factor, some color-neutral particles
may be in the ILC reach. In \reffi{fig:ILC} we show the prediction for
$\sweff$ in
this SPS~1a$'$ inspired scenario as a function of the lighter chargino
mass, $\mcha{1}$. The prediction includes the parametric
uncertainty, $\si^{\rm para-ILC}$, induced by the ILC measurement of $\mt$, 
$\de\mt = 100 \mev$~\cite{mtdet}, and the numerically more
relevant prospective future uncertainty on $\De\al^{(5)}_{\textup{had}}$,
$\de(\De\al^{(5)}_{\textup{had}})=5\times10^{-5}$~\cite{fredl}. 
The MSSM prediction for $\sweff$
is compared with the experimental resolution with GigaZ precision,
$\si^{\rm ILC} = 0.000013$, using for simplicity the current
experimental central value. The SM prediction (with 
$\MHSM = \Mh^{\rm MSSM}$) is also shown, applying again the parametric 
uncertainty $\si^{\rm para-ILC}$.

Despite the fact that no colored SUSY 
particles would be observed at the LHC in this scenario, the ILC with
its high-precision 
measurement of $\sweff$ in the GigaZ mode could resolve indirect effects
of SUSY up to $m_{\tilde\chi^\pm_1} \lsim 500 \gev$. This means that the
high-precision measurements at the ILC with GigaZ option could be
sensitive to indirect effects of SUSY even in a scenario where SUSY
particles have {\em neither \/} been directly detected at the LHC nor the
first phase of the ILC with a centre of mass energy of up to $500 \gev$.


\section{Conclusions}

EWPO provide a very powerful test of the SM
and the MSSM. We have reviewed results for $\MW$ and $\sweff$.
The sensitivity to higher-order effects will drastically improve with
the ILC precision (including the GigaZ/MegaW options) on the EWPO and
$\mt$. This has been illustrated in three examples. 
A precise $\mt$ determination is crucial 

A general scan over
the MSSM parameter space for $\sweff$ and $\mt$ currently does not
prefer the SM or the MSSM over the other. However, the anticipated
GigaZ precision indicates the high 
potential for a significant improvement of the
sensitivity of the electroweak precision tests.
In a second example we have assumed a scenario with very heavy SUSY
particles, outside the reach of the LHC and the first stage of the ILC
with $\sqrt{s} = 500 \gev$. It has been shown that even in such a
scenario the GigaZ precision on $\sweff$ may resolve virtual effects of
SUSY particles, providing a possible hint to the existence of new physics.


\section{Acknowledgments}

Work supported in part by the European Community's Marie-Curie Research
Training Network under contract MRTN-CT-2006-035505
`Tools and Precision Calculations for Physics Discoveries at Colliders'
(HEPTOOLS).


\begin{footnotesize}


\end{footnotesize}